\documentstyle[aps,prb,multicol,epsfig]{revtex}
\draft
\input epsf
\begin{document}
\author{C.~Fusco$^{(1)}$, P.~Gallo$^{(1)}$, A.~Petri$^{(2,3)}$
\footnote{Author to whom correspondence should be addressed. E-mail:
alberto@idac.rm.cnr.it} 
and M.~Rovere$^{(1)}$\\
$^{(1)}$Dipartimento di Fisica and Istituto Nazionale di Fisica della Materia,\\
Universit\`a ``Roma Tre'', Via della Vasca Navale 84, 00146 Roma, Italy\\
$^{(2)}$Consiglio Nazionale delle Ricerche,
Istituto di Acustica ``O. M. Corbino'',\\
Area della Ricerca di Roma Tor Vergata,\\
Via del Fosso del Cavaliere 100, 00133 Roma, Italy\\
$^{(3)}$Dipartimento di Fisica and Istituto Nazionale di Fisica della Materia,\\
Universit\`a ``La Sapienza'', P.le A. Moro 2, 00189 Roma, Italy}
\title{Random sequential adsorption and diffusion of dimers and $k$-mers
on a square lattice} 
\maketitle
\begin{abstract}
We have performed extensive simulations of random sequential adsorption
and diffusion of $k$-mers, up to $k=5$  in two dimensions with particular
attention to the case $k=2$. We focus on the behaviour of the coverage
and of vacancy dynamics as a function of time.
We observe that for $k=2,3$ a complete 
coverage of the lattice is never reached, because of 
the existence of frozen configurations that prevent isolated vacancies in 
the lattice to join. From this result  we argue that complete coverage is
never attained for any value of $k$.
The long time behaviour of the coverage is not mean field and  
non analytic, with $t^{-1/2}$ as leading term.
Long time coverage regimes are independent of 
the initial conditions while strongly depend 
on the diffusion probability and deposition rate and, in particular,
different values of these parameters lead to different  
final values of the coverage. The geometrical complexity of these 
systems is also highlighted through an investigation 
of the vacancy population dynamics.
\end{abstract}
 
\begin{multicols}{2}
\section{Introduction} 
\label{intro}
Random Sequential Adsorption (RSA) of hard particles on lattice is
employed to describe a wide class of irreversible phenomena in physics, 
chemistry and biology.~\cite{evans} The prototype of this process is the 
dimerization of monomers on a chain.~\cite{flory} 
Each monomer occupies initially a 
site of the chain and can react irreversibly with one of its neighbors to form
a dimer: irreversible dimerization takes place sequentially until no
more pairs of neighbor monomers are available.  
According to Flory~\cite{flory} this process
can be viewed as the random sequential adsorption of non-overlapping dimers 
onto a one dimensional lattice: a dimer can be irreversibly deposited if two 
neighbor free sites are available. At a given stage only single isolated sites
will be available and further deposition will be no more allowed.  
At that stage, the density of dimers is finite and takes a non random value in
the limit of infinite system, the so called jamming coverage $\theta_J$.  
This study was followed by the introduction and investigation of a
variety of RSA processes 
involving different objects, lattices
and rules.~\cite{evans} The quantities 
of interest are in general the time behavior of the coverage $\theta(t)$ 
and its jamming value. Exact solutions are available in few cases, typically 
for dimers in one dimension, whereas only numerical or approximate
results exist for more complex objects and in higher 
dimension.~\cite{evans,privman1,privman2,privman3,wang1}
 
In recent years, the class of systems for which RSA yields effective 
description has widened by the observation that in many experimental 
processes particles also diffuse on the lattice after deposition.
Diffusional Relaxation (DR) affects the spatial 
distribution of free sites, 
allowing for further deposition of particles whenever neighbor empty sites 
are created. This produces a different time behavior of $\theta(t)$ and
different values of the asymptotic coverage $\theta(\infty)$.  
The effect of DR on RSA models has been
investigated by many authors (see~\cite{privman2,privman3} and Refs. therein), 
often displaying the emergency of non trivial behaviors both for 
$\theta(t)$ and $\theta(\infty)$.
In the case of dimers on a line, for instance, the introduction of DR produces
a very slow increase of  coverage with time, terminating eventually with the 
complete filling: $\theta(\infty)=1$. 
Exact results~\cite{grynberg1,grynberg2,eisenberg1} and 
numerical simulations~\cite{privman4,nielaba} show that, after a transient 
depending on the initial conditions, $1- \theta(t)$ decreases asymptotically 
as $t^{-1/2}$, and thus an infinite time is needed to reach 
the completely filled state. The exponent $1/2$ is due to the relevance of  
statistical fluctuations in the formation of neighbors free sites, so that any
computation based on mean field treatments is not effective.  
It is also for this reason that 
solutions of RSA problems with Diffusion (RSAD) 
are very rare and usually limited to one dimensional systems 
(see~\cite{privman5} and Refs. therein) and some particular 
cases.~\cite{grynberg1,grynberg2}
%%%%%%%%%%%%%%%%%%%%%%%%%%%%%%%%%%%%%%%%%%%%%%%%%%%%%%%%%%%%%
Some two dimensional models have been recently 
investigated~\cite{privman3,eisenberg2,grigera,eisenberg3}, 
and  features sometimes different from the one 
dimensional case have been observed.  Eisenberg and Baram have found a stable 
asymptotic coverage $\theta(\infty)<1$ for the RSAD of crosses on a 
square lattice~\cite{eisenberg3} whereas such a behavior is not observed in
the case of squares~\cite{eisenberg2}, in which the complete coverage is 
eventually reached for an infinite system, although through coexistence of 
different phases.
The pure RSA of dimers in two dimensions
was first investigated by Nord and Evans~\cite{nord} 
and was later extended also considering anisotropic deposition 
rates.~\cite{deoliveira1} Grigera {\em et al.}~\cite{grigera} computed the 
asymptotic coverage $\theta(\infty)$ in  presence of diffusion
and rotation of dimers and found that it is less than unity. 
They also observed indications for an exponential approach to the limit 
coverage $\theta(\infty)$.
%%%%%%%%%%%%%%%
\begin{figure}[h]
\centering\epsfig{file=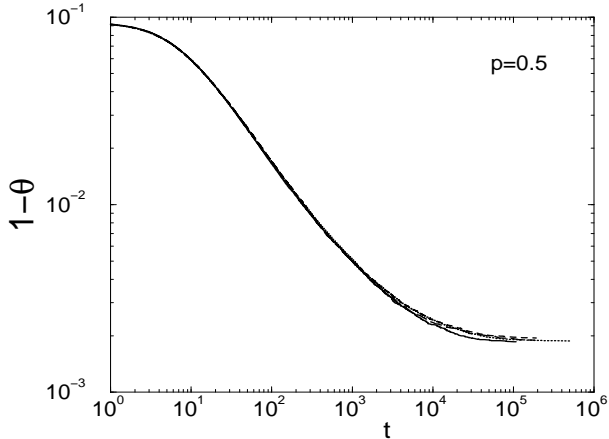, width=8cm, height=6cm}
\caption{Temporal behavior of density of empty sites for different sizes of 
the lattice: $L=40$ (solid line), $L=100$ (dotted line), $L=120$ 
(dashed line), $L=140$ (long dashed line). 
The initial condition is $\theta(0)=\theta_{J}$.}
\protect\label{fig1}
\end{figure}
%%%%%%%%%%%%%

In this paper we present extensive Monte Carlo (MC) simulations 
of the RSAD of linear particles on a square lattice. 
In the systems under consideration hard particles, the $k-$mers, are dropped 
on an initially empty square lattice 
and are allowed to diffuse. Each $k-$mer occupies $k$ 
consecutive sites  and cannot overlap to the others. 
We consider different initial conditions as well as a set of different 
deposition and diffusion rates and probabilities. 
Whereas the former only have influence on  initial and intermediate stages
of the coverage, the latter influence the late stage of dynamics  and  
different values may lead to different final states.
In fact, we show that the value of the final coverage 
depends on these parameter and, as expected, decreases with increasing $k$.
In this paper the focus will be mainly on the case $k=2$, since 
simulations show that its main features are also shared by systems 
with higher
values of $k$, while the time required for simulations rapidly increases
with $k$. For this reason Secs.~\ref{model}-\ref{vacancy} are dedicated
mainly to the RSAD of dimers. 

In Sec.~\ref{model} we define the model and the procedure for the
Montecarlo simulation. We  also test possible algorithm
to save computational time  when simulating such slow relaxation 
processes, and describe the one we adopted. 
Section~\ref{coverage} is devoted to
investigate the dynamical features of the coverage process as well as of
the
final state and of their dependence on the deposition and diffusion rates.
In Sec.~\ref{vacancy}  the results are put in relation with the 
behaviour of the vacancies in the lattice, whose statistical behaviour 
is also studied. 
The $k-$mers RSAD model with $ 2 < k \le 5 $ is addressed in Sec.~\ref{kmers},
where similarities and differences with the case $k=2$ are pointed out.
Finally sec.~\ref{conclusions} is devoted to 
a summary of our main findings and to some general conclusions.

\section{Model and Computational Details}
\label{model}

\subsection{RSAD dynamics}
The process that we have more extensively
investigated in this paper consists in the deposition and 
diffusion of dimers on a square lattice containing $N=L\times L$ sites with 
periodic boundary conditions. Each dimer occupies two nearest neighbor sites,
or equivalently a bond, and cannot overlap with other dimers. For
$t > 0$ it is free to diffuse along the direction of its axis towards empty 
sites. We have assumed deposition to occur with equal probability on 
horizontal and vertical bonds and we have considered two possible different 
initial conditions: $a)$ jamming coverage, i.e. $\theta(0)=\theta_J$ and 
$b)$ empty lattice, i.e. $\theta(0)=0$.  In the case $a)$ the initial stage
consists of pure RSA. Dimers are deposited on the lattice until no more free 
pair of sites is available and the jamming state is reached.  
The jamming coverage for this kind of model is known from previous 
work~\cite{deoliveira1} to be $\theta_J=0.90654$. It is reached
exponentially and is higher than the corresponding one in one dimension. At
this stage deposited dimers are allowed to diffuse and more dimers can be 
possibly added when suitable space is made available.  
In case $b)$ the 
deposited dimers can move from the beginning,
so that diffusion in the initial stages largely dominates the deposition 
process, giving rise to a large amount of vacancies.  

\begin{table} 
\caption{\label{tab1} Coefficients of fit Eq.~(\ref{fit1}) for 
$\theta(0)=\theta_{J}$. The curve for $p=0.5$ can be adequately fitted only
retaining the terms up to the third order.} 
\begin{tabular}{|c|c|c|c|c|c|}
\hline
$p$       &       $r_{0}$     &       $r_{1}$      &        $r_{2}$     &        $r_{3}$   &    $r_{4}$   \\
\hline
  
0.2       &       0.00150     &       0.040        &        2.167       &        7.447     &    7.456     \\ 
       
0.5       &       0.00110     &       0.104        &        0.607       &        -1.105    &              \\
 
0.9       &       0.00120     &       0.252        &        0.702       &        -3.983    &    4.799     \\ 
\hline
\end{tabular} 
\end{table}

\begin{table} 
\caption{\label{tab2} Coefficients of fit Eq.~(\ref{fit1}) for 
$\theta(0)=0$.}
\vspace{0.5cm}
\begin{tabular}{|c|c|c|c|c|c|} 
\hline
$p$       &       $r_{0}$     &       $r_{1}$      &        $r_{2}$     &        $r_{3}$   &    $r_{4}$   \\
\hline
0.2       &       0.0016      &       0.0628       &        1.358       &
  2.152     &    -4.836    \\

0.5       &       0.00084     &       0.1239       &        0.4223      &
  0.2943    &    -0.3776   \\

0.9       &       -0.0014     &       0.3797       &        -0.6506     &
  1.021     &     -0.4146  \\
\hline
\end{tabular}
\end{table}           
\begin{figure}
\centering\epsfig{file=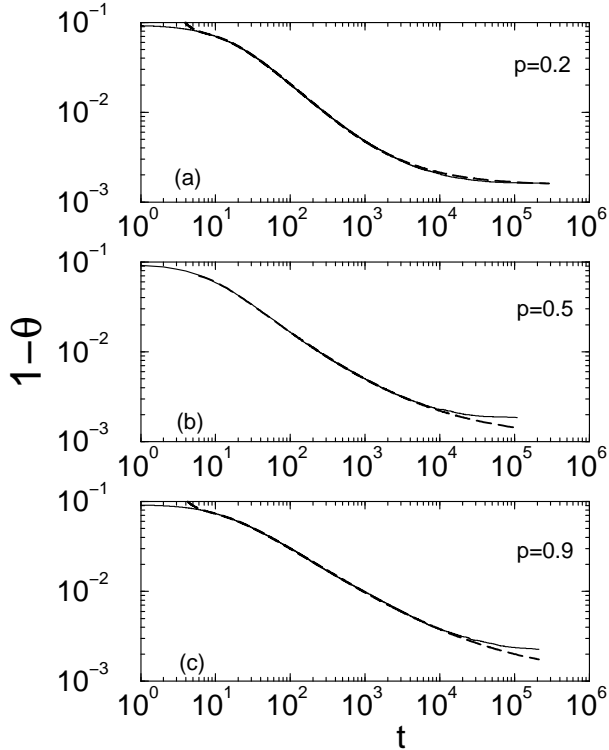, width=8cm, height=10cm}
\caption{Temporal behavior of density of empty sites (solid lines) for $p=0.2$
(a), $p=0.5$ (b) and $p=0.9$ (c). The long dashed lines are fits to the data 
according to Eq.~(\ref{fit1}). The initial condition is $\theta(0)=\theta_{J}$
and the size of the system is $L=100$.}
\protect\label{fig2}
\end{figure}

We have computed the time behavior of the coverage and of the 
abundance of different 
kinds of vacancies  in the above processes by means of MC 
simulations. At each time step deposition of dimers is attempted with 
probability $p$ and diffusion with probability $1-p$. 
Both directions of diffusion are equiprobable. Deposition happens
at rate $R$, i.e. $pRN$ dimers are tried to be deposited on the average in a 
MC step and a unitary jumping rate $d=1$ follows for diffusion by 
assuming unitary 
lattice spacing and simulation (micro) time discretization $\delta t =1/N$. 

\subsection{Saving time algorithms}
The above dynamics is extremely time consuming to simulate, mainly in the late
stage of the process, since due to the exclusion rules most of diffusion and 
deposition attempts fail. A large amount of computational time can however be 
saved and dynamics sped up by suitable algorithms that maximize the number of 
successful events.  We have adopted the strategy of storing in memory the
dimers that can successfully diffuse along a given direction, 
$n_{difH}$ and $n_{difV}$ and the bonds available for deposition $n_{depH}$ 
and $n_{depV}$ respectively for the horizontal and vertical directions. 
%%%%%%%%%%%%%%%%
\begin{figure}
\centering\epsfig{file=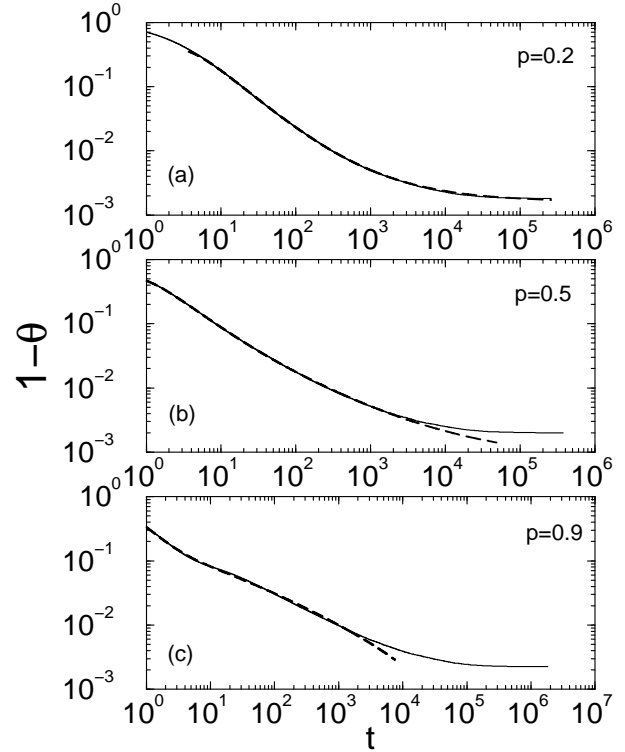, width=8cm, height=10cm}
\vspace{0.9cm}
\caption{Temporal behavior of density of empty sites (solid lines) for $p=0.2$
(a), $p=0.5$ (b) and $p=0.9$ (c). The long dashed lines are fits to the data 
according to Eq.~(\ref{fit1}). The initial condition is $\theta(0)=0$
and the size of the system is $L=100$.}
\protect\label{fig3}
\end{figure}
%%%%%%%%%%%
By accounting for the different probabilities and rates at
which these actions may take place, it turns out that on the average a 
successful event occurs with probability  
\begin{equation}
\label{speedup} 
p_{succ} = p \left(\frac{n_{depH}+n_{depV}}{2N}\right)+\frac{1-p}{2}
\left(\frac{n_{difH} +n_{difV}}{2N}\right).
\end{equation}
Since the law of large numbers is expected to hold (at least for 
system size large enough), at each step of the simulation the MC time 
is increased by the 
amount
\begin{equation}
\label{time}  
\Delta t=\frac{1}{p_{succ}}. 
\end{equation}
We adopted 
the above method that produces a time saving of about 
one order of magnitude with
respect to the usual MC simulation 
with equivalent results.

As a further check we have also tested the method adopted by Gan and 
Wang~\cite{gan}, which is based on the probability distribution of time lags 
$\Delta t$ between two consecutive successful events
\begin{equation}
\label{distribution} 
P(\Delta t=m)= p_{succ}(1- p_{succ})^{m-1}
\end{equation}
and also yields results equivalent to those of usual MC.  
The identity of results confirms  the reliability of both of these
methods; however time saving is much more efficient via Eq.~(\ref{speedup}), 
since in fact this latter method corresponds to an average of the former over 
a large number of successful events. Both of them require additional memory 
occupation for the storage of dimers and bonds that may change their state, 
but this does not seriously hinder simulations of large systems if compared 
with required computational time. 

\begin{figure}
\centering\epsfig{file=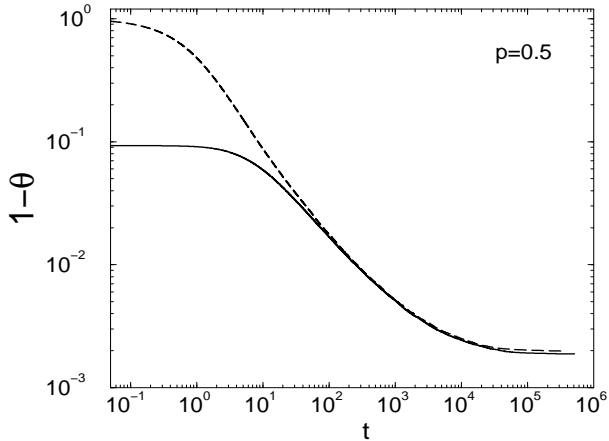, width=8cm, height=6cm}
\caption{Comparison of the temporal behavior of density of empty sites 
for an initial jammed lattice (solid line) and an initial empty lattice 
(long dashed line), for $p=0.5$ The size of the system is $L=100$.}
\protect\label{fig4}
\end{figure}

\section{Time coverage}
\label{coverage}

\subsection{Asymptotic coverage} 
Mean field approximations are known to work very badly for RSAD of small 
particles in one dimension. Diffusional dynamics of dimers along the lattice
can be described in terms of mobile vacancies, 
that annihilate through the deposition of a dimer when they meet on 
neighbor sites. As already mentioned, both exact solutions and
numerical simulations show that the process of vacancies annihilation is 
dominated by random fluctuations of density rather than by average
values, thus leading asymptotically to a coverage described by a random walk 
process, $\theta(\infty)-\theta(t) \simeq t^{-1/2}$, opposed to the $t^{-1}$ 
coverage estimated by mean field. 
Random walk process with regular exponent $0.5$ concerns 
the dynamics of vacancies, since
particles suffer an anomalous random walk with exponent $\le 0.5 $, as
discussed in Ref.~\cite{fusco}.
 
Mean field is generally speaking expected to work better in 
higher dimensions. However studies on RSAD of different objects in $d=2$ 
found that fluctuations of isolated vacancies still dominate
relaxational dynamics.~\cite{eisenberg2,wang2,wang3} 
Investigations of Eisenberg and Baram~\cite{eisenberg2} on RSAD of monomers 
particles with nearest neighbors and next nearest neighbors 
exclusion have shown that $\theta(\infty)-\theta(t) \simeq t^{-1/2}$ to the
leading order in the asymptotic regime. 
Systems made by  superposition  of several
one dimensional lattices to form  multilayer structures, that allow for both
intra-layer and inter-layer diffusion, have been found to 
behave in a similar fashion.~\cite{deoliveira2}

Our simulations show that 
also the RSAD of dimers in $d=2$ displays a regime in which  
$t^{-1/2}$ dominates,
thus implying the recombination of one-site vacancies to be
the main 
process, in analogy with the  $d=1$ case (we will turn to this point in 
Sec.~\ref{vacancy}). This result has shown to be 
independent from the parameter values and initial conditions chosen for the  
simulation. Nevertheless in $d=2$ there is a stage of the covering process  
at which the dynamics definitely changes. At that stage we observe that
the annihilation rate slows down and 
correspondingly the filling slowly stops. The system is prevented to  
reach the complete coverage by the formation of isolated vacancies that are 
trapped at one site or along closed loops. 
As firstly pointed out by Grigera {\em et al.}~\cite{grigera} 
this seems to be a genuine property of the system, 
since independent of the lattice size.  
The results displayed in Fig.~\ref{fig1}, where systems of 
different sizes approach the same asymptotic coverage  
$\theta(\infty) \simeq 0.998$, 
support the idea of a dynamically jammed state related to local
configurations of the system.
We find, in agreement with the authors in~\cite{grigera}, that finite-size 
effects are reasonably negligible for $L\ge 40$. 

In Fig.~\ref{fig2} we show the coverage behavior
for different choices of $p$.   		
Different values 
of $\theta(\infty)$ are reached, implying that 
different choices of $p$ and $R$ can
lead to different final coverages
and can therefore be regarded as different quench rates of an 
initially annealed system.

\begin{figure}
\centering\epsfig{file=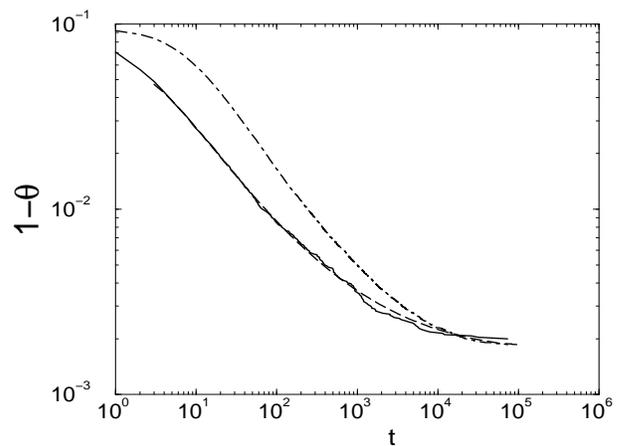, width=8cm, height=6cm}
\caption{Temporal behavior of density of empty sites for infinite deposition
rate (solid line). The long dashed line is a fit to the data according to 
Eq.~(\ref{fit2}). 
For comparison we also show the curve for $p=0.5$ (dot-dashed line). 
The initial condition is $\theta(0)=0$ and the size of the system is $L=100$.}
\protect\label{fig5}
\end{figure}
\begin{figure}
\centering\epsfig{file=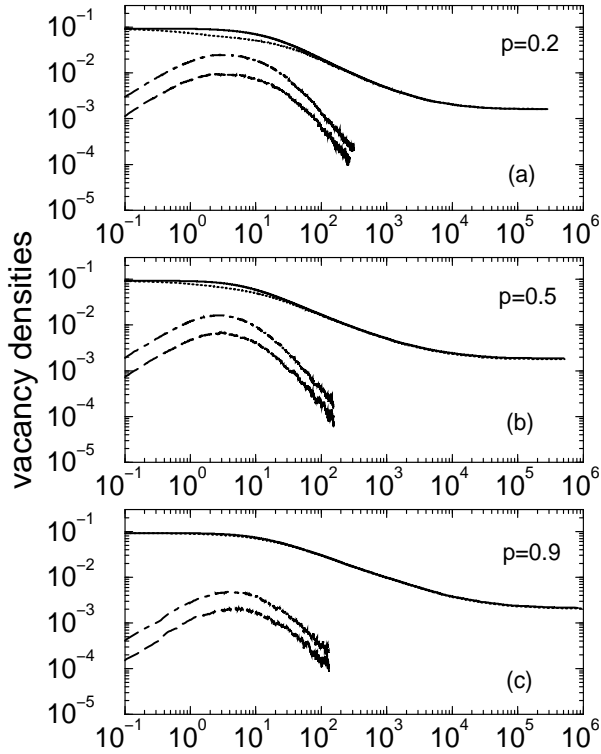, width=8cm, height=10cm}
\caption{Temporal behavior of different kinds of vacancies for $p=0.2$ (a),
$p=0.5$ (b) and $p=0.9$ (c), for $\theta(0)=\theta_{J}$: the solid line is
the total density of empty sites, the dotted line is the density of one-site 
vacancies, the dashed line is the density of sites formed by two-site 
vacancies and the dot-dashed line is the density of sites formed by non 
one-site vacancies. The size of the system is $L=100$.}
\protect\label{fig6}
\end{figure}

\subsection{System evolution} 
Fig.~\ref{fig2} puts into evidence a somewhat counterintuitive feature, i.e. 
that faster coverages are obtained at lower deposition probabilities, 
showing that to this aim the process of vacancies generation is more important
than  deposition attempts themselves.
In order to understand this behavior it 
is worth to consider the different processes that contribute to 
increase the system coverage. This can be done by
describing the dynamics of the system
on a wide time scale by a series of the form
\begin{equation}
\label{fit1}
1-\theta(t)=\sum_{i=0}^{\infty}r_it^{-i/2}
\end{equation}
A fit of this curve
to the results of  numerical simulations shows that 
the terms beyond the fourth in the series can be neglected 
(cfr.~Table~\ref{tab1}) and is displayed in Fig.~\ref{fig2}.
As expected, the effect of an increased 
diffusion is to decorrelate the system, 
and thus to produce  a faster coverage,
as shown by higher weights for the $t^{-1}$ term.  
The coverage depends on the initial condition only at
early times.
%%%%%%%%%%%%%%%%%%%%%%%%%%%%%%%%%%%%%%%%%%%%%%%%%
\begin{figure}
\centering\epsfig{file=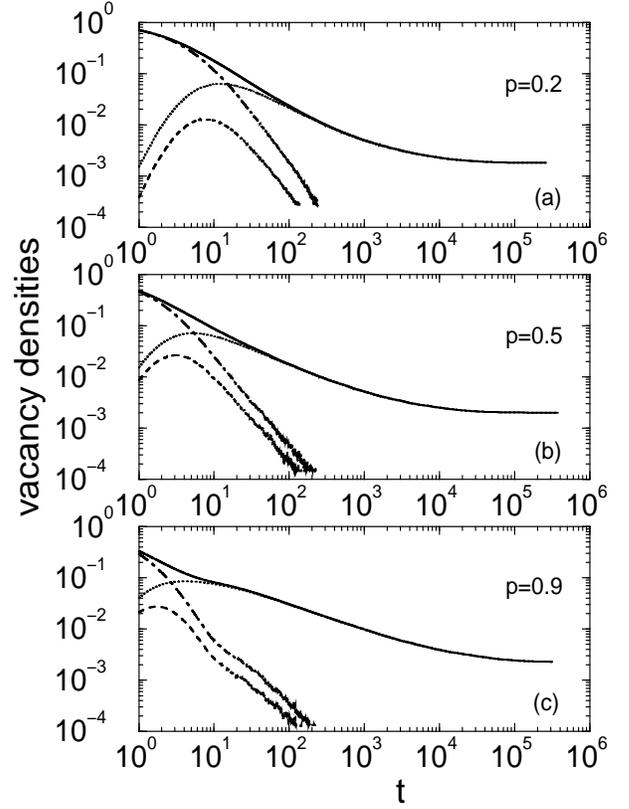, width=8cm, height=10cm}
\vspace{1cm}
\caption{Temporal behavior of different kinds of vacancies for $p=0.2$ (a),
$p=0.5$ (b) and $p=0.9$ (c), for $\theta(0)=0$: the solid line is
the total density of empty sites, the dotted line is the density of one-site 
vacancies, the dashed line is the density of sites formed by two-site 
vacancies and the dot-dashed line is the density of sites formed by non 
one-site vacancies. The size of the system is $L=100$.}
\protect\label{fig7}
\end{figure}
%%%%%%%%%%%%%%%%%%%%%%%%%%%%%%%%%%%%%%%%%%%%%%%%%
Fig.~\ref{fig3} displays the time coverage when $\theta(0)=0$. 
After an initial stage ($1<t<10^{2}$) in which deposition attempts 
dominate and the density progressively increases, 
the system approaches the same regime as for  $\theta(0)=\theta_{J}$. 
%%%%%%%%%%%%%%%%%%%%%%%%%%%%%%%%%
\begin{table}
\caption{\label{tab3} Coefficients of fit Eq.~(\ref{fit3}) for 
$\theta(0)=\theta_{J}$.}
\vspace{0.5cm}
\begin{tabular}{|c |c |c |c |c| c|}  
\hline
$p$       &       $x_{0}$     &       $x_{1}$      &        $x_{2}$     &        $x_{3}$   &    $x_{4}$   \\
\hline

0.2       &       0.00131     &       0.093        &        0.834       &        2.932     &    2.833     \\ 
       
0.5       &       -0.00158    &       0.174        &        0.602       &        -2.916    &    3.016     \\
 
0.9       &       0.00117     &       0.249        &        0.575       &        -4.801    &    6.437     \\
\hline
\end{tabular}
\end{table}
%%%%%%%%%%%%%%
\begin{table}
\caption{\label{tab4} Coefficients of fit Eq.~(\ref{fit4}) for 
$\theta(0)=\theta_{J}$.}
\vspace{0.5cm}
\begin{tabular}{|c |c |c |c |c|}  
\hline
$p$       &       $y_{0}$     &       $y_{1}$      &        $y_{2}$     &        $y_{3}$        \\
        \hline

0.2       &       0.00222     &       0.0509       &        0.371       &        -0.446         \\ 
       
0.5       &       0.00306     &       -0.0946      &        0.925       &        -1.484         \\
 
0.9       &       -0.00010    &       -0.0033      &        0.0896      &        -0.1268        \\
\hline
\end{tabular}
\end{table}
%%%%%%%%%%%%%%%%%%%

\begin{figure}
\centering\epsfig{file=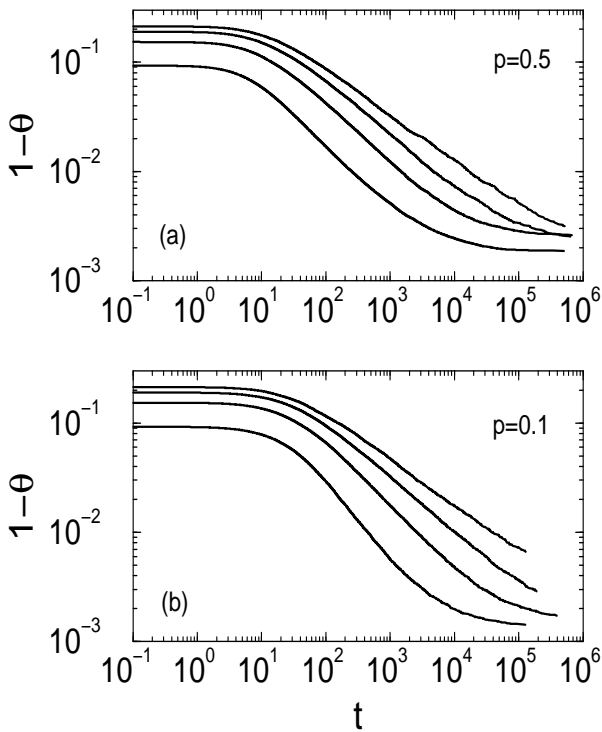, width=8cm, height=10cm}
\caption{Temporal behavior of empty sites density for $p=0.5$ (a) and $p=0.1$
(b), at different values of particle length $k$: $k=2,3,4,5$ (from bottom to 
top in each graph). The initial condition is $\theta(0)=\theta_{J}$ and the 
size of the system is $L=100$.}
\protect\label{fig8}
\end{figure}
A comparison of the two cases is made in Fig.~\ref{fig4}. 
This time behavior can also be well described by curves 
of the form of Eq.~(\ref{fit1}), obtaining 
the coefficients of Table~\ref{tab2}.  
Coefficients relevant to the late stage process exhibit a
similar trend with respect to those of Table~\ref{tab1}.            

Another interesting situation, also often investigated in literature,  
is the Diffusion Limited Deposition (DLD). In this 
case $p=1$ and $R=\infty$, 
corresponding to the filling of two-site vacancies with dimers  as soon as 
they are created by diffusion. 
In this case only one-site vacancies exist, 
making the coverage process much more effective with respect to the case 
of finite $R$, as evidenced in Fig.~\ref{fig5}. 

On the other hand the asymptotic coverage is higher for finite deposition 
rate, showing that the larger diffusion allowed becomes important at very 
high coverage values, since in that case the long time frozen configurations
are more easily removable. Thus, in analogy with Figs.~\ref{fig2} and 
\ref{fig3}, different 
asymptotic coverages may be reached. Also for DLD a superposition of non
analytic 
powers of time fits the curve quite well in all the dynamical range:
\begin{equation}
\label{fit2}
1-\theta(t) \simeq b_{0}+b_{1}t^{-1/2}+b_{2}t^{-1}+b_{3}t^{-3/2},
\end{equation} 
where $b_{0}\simeq 0.0016$, $b_{1}\simeq 0.060$, $b_{2}\simeq 0.112$ and 
$b_{3}\simeq -0.137$ as shown in Fig.~\ref{fig5}. 

\section{Vacancy dynamics} 
\label{vacancy}
As mentioned above, RSAD can be described in terms of reaction and diffusion 
of vacancies on the lattice.
In the general case of $k-$mers with diffusion coefficient $D$ and deposition 
rate $R$, one deposition corresponds to the annihilation of
$k$ vacancies on contiguous sites. 

If one adopts this point of view the system behavior appears to be ruled by 
the dynamics of vacancies.
In order to give a full account of this dynamics, knowledge of vacancies 
correlation functions is necessary. Nevertheless a qualitative understanding 
of some of the main dynamical features can be obtained also through the 
investigation of the sole vacancy population.

\subsection{Initial jammed state}
Fig.~\ref{fig6} compares the coverage behavior with the change in time of  
vacancy population for different values of $p$. Solid and dotted lines 
represent respectively the total fraction of empty sites, $1-\theta$, 
and of one-site vacancies, $\rho_1$. At early and intermediate 
times these two quantities
display some differences, more visible at lower values of $p$ where formation 
of many-site vacancies is enhanced by diffusion. For a jammed initial state, 
an increase in two-site vacancies takes place in
the early stage of the process and, for low values of $p$, 
can last for several decades both in density
($\rho_2$) and time. 
In fact, two dimensional geometry allows for the formation of many 
different kinds of voids; besides vertical and horizontal two-site vacancies, 
there can be $L$ many-site vacancies with different orientations, 
square vacancies 
and so on. 
%%%%%%%%%%%%%%%
\begin{table}
\caption{\label{tab5} Jamming densities of $k-$mers.}
\begin{tabular}{|l|l|}
$k$  & $\theta_{J}$\\
\hline
3 &  0.8470 \\
4 &  0.8103  \\ 
5 &  0.7877 \\   
\end{tabular}
\end{table}
%%%%%%%%%%%%%%555
\begin{table}
\caption{\label{tab6} Coefficients of fit Eq.~(\ref{fit1}) for $k-$mers when
$\theta(0)=\theta_{J}$.}
\vspace{0.5cm}
\begin{tabular}{|c |c |c |c |c|c|}  
\hline
$k$         &       $r_{0}$      &     $r_{1}$     &     $r_{2}$     &     $r_{3}$     &      $r_{4}$    \\
\hline
2 ($p=0.5$) &       0.00110      &     0.104       &     0.607       &     -1.105      &                 \\

3 ($p=0.5$) &       0.00075      &     0.329       &     1.562       &     -8.275      &      11.51      \\

4 ($p=0.5$) &       0.00036      &     0.682       &     0.399       &     -6.709      &      10.89      \\

5 ($p=0.5$) &       0.00228      &     1.013       &     -1.774      &     0.8537      &      0.373      \\
\hline 
2 ($p=0.1$) &       0.00126      &     0.027       &     4.174       &     -19.30      &      26.36      \\ 

3 ($p=0.1$) &       0.00036      &     0.384       &     6.483       &     -47.35      &      95.37      \\  

4 ($p=0.1$) &       -0.00031     &     1.033       &     0.474       &     -17.13      &      37.44      \\

5 ($p=0.1$) &       0.00075      &     1.663       &     -6.330      &     17.38       &      -20.80     \\
\end{tabular}
\end{table}
%%%%%%%%%%%%%5
\begin{figure}
\centering\epsfig{file=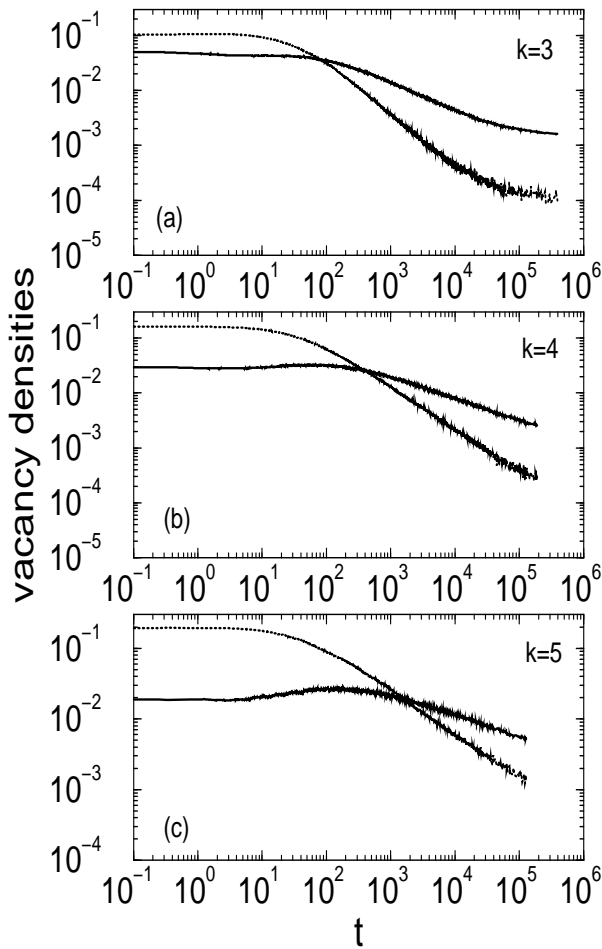, width=8cm, height=12cm}
\vspace{0.9cm}
\caption{Temporal behavior of density of one-site vacancies (solid lines) and
of sites formed by non one-site vacancies (dotted lines) for $p=0.1$ and 
$k=3$ (a), $k=4$ (b), $k=5$ (c). The initial condition is 
$\theta(0)=\theta_{J}$ and the size of the system is $L=100$.}
\protect\label{fig9}
\end{figure}

A comparison of $\rho_2$ with the overall non-one-site vacancy 
density, $1-\rho_1-\theta$, is shown in
Fig.~\ref{fig6}. We observe that the abundance of the latter 
does not vanish for 
large times, reflecting  the  configurational richness of the 
two-dimensional process with respect to the unidimensional one  
where  $\rho_{2}$ and $1-\rho_{1}-\theta$ coincide asymptotically.

If one expands $\rho_2$ and $1-\rho_1-\theta$ in powers of $t^{-1/2}$ 
one finds,
as for the coverage, that
larger contributions from mean field terms appear at lower 
$p$ values in the asymptotic  expansions: 
\begin{equation}
\label{fit3}
\rho_2(t)=\sum_{i=0}^{\infty}x_it^{-i/2},      
\end{equation}

\begin{equation} 
\label{fit4} 
1-\rho_1(t)-\theta(t)=\sum_{i=0}^{\infty}y_it^{-i/2}.
\end{equation}   

The coefficients of the best fit of Eqs.~(\ref{fit3}) and (\ref{fit4}) to the
numerically determined curves for $\rho_{2}$ and $1-\rho_{1}-\theta$ are 
reported respectively in Table~\ref{tab3} and Table~\ref{tab4}.

\subsection{Initial empty lattice}
An initially empty lattice produces early and intermediate stages 
in vacancy dynamics different from the jammed case.
Whereas the late stage is still dominated by recombination of 
one-site vacancies, an initial stage 
exists in which the abundance of this kind of 
defects increases monotonically
up to reach about the same density of many-site vacancies, as displayed 
in  Fig.~\ref{fig7}. 
On the other hand, right at that stage the latter's population starts to
be less than that of one-site vacancies. 
Interestingly single empty sites rule the dynamics for $t>10^{2}$, that
is, as already explained, the same time range in which  density 
no longer depends on initial conditions (cfr. Fig.~\ref{fig4}). 
It can be noted that two-site vacancies grow at pace with one-site
ones up to a maximum, where they start to decay at the same rate of larger 
ones. The location of this maximum as well as the intersection between
$2\rho_2$ and $1-\rho_{1}-\theta$ tend to move to early times when $p$
increases, suggesting that enhanced deposition rate destroys 
many-site vacancies more efficiently. The functional forms that fit well 
$\rho_{2}$ and
$1-\rho_{1}-\theta$ are the same as those for the initial jammed 
lattice case (cfr. Eqs.~(\ref{fit3}) and (\ref{fit4})).

\section{RSAD of $k-$mers}
\label{kmers}
We now address the problem of RSAD of $k-$mers on a square lattice,
which is a 
generalization of the model described in Sec.~\ref{model} 
that corresponds to $k=2$. 
Since for increasing $k$ the simulations of the dynamics 
becomes more and more time consuming and finite-size effects  more 
important, we 
concentrate on low values of $k$, namely $k=3,4,5$.

We have simulated the RSAD dynamics always starting  with an initial jammed 
state and then switching on the diffusive part of the process.
We checked that the jamming densities obtained as initial conditions 
after the pure RSA process were in agreement with the values found earlier 
by Bonnier {\em et al}.~\cite{bonnier} It turns out that there were no 
discrepancies with their results within the numerical precision.
The 
jamming densities obtained are summarized in the Table~\ref{tab5}.

In Fig.~\ref{fig8} we compare the temporal evolution of the $k-$mers density 
for two values of $p$. 
%%%%%%%%%%%%%%%%%55
\begin{table}
\caption{\label{tab7} Crossover time for one site vacancy domination.} 
\begin{tabular}{|r|r|}
$k$ &   $t_{c}$\\
\hline
3 &  80 \\
4 &  400 \\ 
5 &  1680 \\   
\end{tabular}
\end{table}
%%%%%%%%%%%%%5
The major aspect to underline is the very sluggish 
dynamical behavior for increasing $k$, which is a common feature with the one 
dimensional case: the probability for $k$ vacancies to meet and annihilate
decreases when $k$ gets larger. 
It can be noted that, as for dimers, a
dynamical saturation at long times exists, although our data are less clear
for $k=5$, since the time at which this plateau 
sets in is an increasing function
of $k$. Such a saturation is absent in $d=1$, where the density follows a 
single power law for long times also for larger values of $k$.\cite{nielaba}  
It appears therefore to be a feature typical  of the two dimensional geometry 
and  independent of the particle size,  related to the complexity 
of the configurational space accessible to the system. 
A proof of that  is that  
the curves in Fig.~\ref{fig8} can be fitted by the same functional 
form adopted for dimers (Eq.~(\ref{fit1})) with the parameters of 
Table~\ref{tab6}.
It is seen  that the $t^{-1/2}$ term is more important at larger $k$,     
in agreement with a slow down in the coverage process.

Arguments based on diffusive reactions~\cite{kang} have shown in the past that 
a critical dimension $d_{c}=2/(k-1)$ should exist above which fluctuations
would be negligible and mean field behavior ($1-\theta\sim t^{-1/(k-1)}$)
would dominate. This would imply that for $d=2$ the mean field law should 
work for $k\ge 3$; nevertheless we have found no evidence 
of that, at least for 
$k=3$. In order to draw such a conclusion
for larger $k$, longer runs would 
be required, since a change in the dynamical behavior 
could occur at very large times. 

Finally we have compared the data for the density 
of one-site vacancies and of many-site vacancies, for $k=3,4,5$
and $p=0.1$ (Fig.~\ref {fig9}). 
At variance with the case $k=2$, many-site vacancies tend now to prevail for 
a long time and  
disappear more slowly for larger $k$; the time behavior of their density
can be well approximated by a single power law:
\begin{equation}
1-\rho_{1}(t)-\theta(t) \sim t^{-\beta_{k}},
\end{equation}
with $\beta_{3}=0.94$, $\beta_{4}=0.75$ and $\beta_{5}=0.61$.
Moreover, the crossover time $t_{c}$ after which one-site vacancies dominate 
increases with $k$, as shown in table~\ref{tab7} of $t_c$ values,
where we note 
that for higher $k$ values diffusive regime sets in at later time.

\section{Summary and conclusions}
\label{conclusions}   
We studied irreversible deposition and diffusion of dimers and $k$-mers on a 
square lattice, using a driven algorithm which allowed to decrease the 
CPU time needed for Monte Carlo simulations. Both time 
coverage behavior and vacancy dynamics
have been analyzed in detail as a function of initial conditions and deposition
rate and probability. A wider variety of dynamical regimes with 
respect to the one 
dimensional case has been observed, reflecting the  complexity of
the configurational space accessible to dimers in $d=2$ with respect to
$d=1$. In particular, the
existence of dynamically frozen configurations, that only depend on the local
environment, slows down the asymptotic dynamics and does not allow the system 
to reach the complete coverage. 
Different choices of deposition and diffusion rates and probabilities
can lead to different final values of the asymptotic coverage.
Simulation shows that
the behavior of coverage $\theta$ strongly deviates from the mean field
prediction.
Deviations from mean field are less pronounced when the 
diffusion probability is higher. 
Furthermore $\theta$ is observed to depend
on the initial conditions only in the early stage of the dynamics. Starting
from the jammed state correlations are higher and the dynamics evolves 
more slowly. Vacancy dynamics is dominated not only by single-site vacancies, 
as in $d=1$, but also by larger vacancies. 

It was also shown that the dynamical behavior is severely slowed down with 
the increase of the length $k$ of the particles. We do not have
evidences of a mean field behavior, in spite of the fact that it would be 
expected to hold for $k\ge 3$.
We have also pointed out that the dimensionality plays
an important role in  
reaction-diffusion models with the result of a rich dynamical 
behavior in the investigated quantities. 

As an open possibility 
for the future, we think that in order 
to better understand the dynamics
of these models it would be useful to vary the kind of relaxation mechanisms,
investigating the possibility of occurrence of collective effects in $d=2$ and 
$d=3$.

\section*{Acknowledgments}  
C.~F. would like to thank C.K.~Gan and J.S.~Wang for having received the code
of their algorithm. A.~P. enjoyed many useful discussions with  
M.J.~de~Oliveira.\\

%%%%%%%%%%%%%%%%%%%%%%%%%%%%%%%%%%%%%%%%%%%%%%%%%%%%%%%%%%%%%%%%%%%%%%%%%%
\newpage

\end{multicols}


\begin{thebibliography}{99}

\bibitem{evans} J.V. Evans, Rev. Mod. Phys. {\bf 65}, 1281 (1993).
\bibitem{flory} P.J. Flory, J. Am. Chem. Soc. {\bf 61}, 1518 (1939). 
\bibitem{privman1} V. Privman, Trends in Statistical Physics 1, p. 89-95
(Council for Scientific Information, India, 1994). 
\bibitem{privman2} V. Privman, Ann. Rev. Comp. Phys., Vol. {\bf III}, 
p. 177-193, ed. D. Stauffer (World Scientific, Sigapore, 1995).
\bibitem{privman3} V. Privman, Colloids and Surfaces A {\bf 165}, 231 (2000).
\bibitem{wang1} J.S. Wang, Colloids and Surfaces A {\bf 165}, 325 (2000).
\bibitem{grynberg1} M.D. Grynberg, T.J. Newman and R.B. Stinchcombe,
Phys. Rev. E {\bf 50}, 957 (1994).
\bibitem{grynberg2} M.D. Grynberg and R.B. Stinchcombe,
Phys. Rev. Lett. {\bf 74}, 1242 (1995).
\bibitem{eisenberg1} E. Eisenberg and A. Baram, J. Phys. A {\bf 30}, L271 
(1997).
\bibitem{privman4} V. Privman and P. Nielaba, Europhys. Lett. {\bf 18}, 673
(1992).
\bibitem{nielaba} P. Nielaba and V. Privman, Mod. Phys. Lett. B {\bf 6}, 533
(1992).
\bibitem{privman5} Nonequilibrium Statistical Mechanics in One
Dimension, ed. V. Privman (Cambridge University Press, 1997).
\bibitem{eisenberg2} E. Eisenberg  and A. Baram, Europhys. Lett. {\bf 44}, 
168 (1998) and ref. therein.
\bibitem{grigera} S.A. Grigera, T.S. Grigera and  J.R. Grigera, Phys. Lett. A
{\bf 55}, 124 (1997). 
\bibitem{eisenberg3} E. Eisenberg and A. Baram, J. Phys. A {\bf 33}, 1729 
(2000).
\bibitem{nord} R.S. Nord and J.W. Evans, J. Chem. Phys. {\bf 82}, 2795 (1985).
\bibitem{deoliveira1} M.J. de Oliveira, T. Tom\'e and R. Dickman, 
Phys. Rev. A {\bf 46}, 6294 (1992).
\bibitem{gan} C.K. Gan and J.S. Wang, Phys. Rev.  E {\bf 55}, 107 (1997).
\bibitem{fusco} C. Fusco, P. Gallo, A. Petri and M. Rovere, 
{\em Stretched exponential relaxation in a diffusive model for granular
compaction}, submitted for publication (2001).
\bibitem{wang2} J.S. Wang, P. Nielaba and V. Privman, Mod. Phys. Lett. B 
{\bf 7}, 189 (1993).
\bibitem{wang3} J.S. Wang, P. Nielaba and V. Privman, Physica A {\bf 199}, 527
(1993).
\bibitem{deoliveira2} M.J. de Oliveira and A. Petri,  J. Phys. A {\bf 31}, 
L425 (1998).
\bibitem{bonnier} B. Bonnier, M. Hontebeyrie, Y. Leroyer, C. Meyers and
E. Pommiers, Phys. Rev. E {\bf 49}, 305 (1994).
\bibitem{kang} K. Kang, P. Meakin, J.H. Oh e S. Redner, J. Phys. A {\bf 17}, 
L665 (1984).

\end{thebibliography}
\end{document}